\documentclass[prd,twocolumn,amsmath,amssymb,nofootinbib,preprintnumbers,balancelastpage]{revtex4}
\usepackage{graphicx}
\usepackage{hyperref}
\hypersetup{
    pdfnewwindow=true,     
    colorlinks=true,       
    linkcolor=black,       
    citecolor=blue,        
    filecolor=blue,        
    urlcolor=blue          
}
\begin{document}
\title{ \Large{\bf{Left--Right Symmetric Theory with Light Sterile Neutrinos}}}
\author{Michael Duerr, Pavel Fileviez P\'erez, Manfred Lindner}
\affiliation{\vspace{0.15cm} \\ Particle and Astro-Particle Physics Division \\
Max Planck Institute for Nuclear Physics {\rm{(MPIK)}} \\
Saupfercheckweg 1, 69117 Heidelberg, Germany}
\preprint{}
\begin{abstract}
A simple theoretical framework for the spontaneous breaking of parity, and baryon and lepton numbers is proposed.
In this context, the baryon and lepton numbers are independent local gauge symmetries, while parity is defined making use of the left--right symmetry. 
We show that in the minimal model the new leptoquark fields needed to define an anomaly-free theory also generate neutrino masses through the type III seesaw mechanism. 
The spectrum of neutrinos and some phenomenological aspects are discussed. This theory predicts the possible existence of two light sterile neutrinos.
\end{abstract}
\maketitle
\section{Introduction}
The existence of massive neutrinos in nature has motivated an enormous number 
of theoretical studies in order to understand the origin of their masses. 
There are three simple mechanisms for the generation of neutrino masses at tree level.  In the type I seesaw mechanism one adds at least two fermionic singlets, $N \sim (1,1,0)$~\cite{TypeI-1,TypeI-5,TypeI-3,TypeI-4,TypeI-2}, while in the type II seesaw mechanism the Standard Model (SM) 
particle content is extended by an extra scalar triplet, $\Delta \sim (1,3,1)$~\cite{TypeII-1,TypeII-2,TypeII-3,TypeII-4,TypeII-5}. 
Another option is to use the type III seesaw mechanism where at least two 
extra fermionic triplets with zero weak hypercharge, $\rho \sim (1,3,0)$~\cite{TypeIII}, are added. 
See Refs.~\cite{Ma,Colored} for other mechanisms. All these mechanisms can be 
realized in the context of left--right symmetric theories or grand unified theories 
where the seesaw scale might be related to other physical scales.

Left--right symmetric theories~\cite{Pati-Salam,Pati-Mohapatra,Senjanovic-Mohapatra,GoranLR} are considered to be very appealing 
candidates for physics beyond the SM. These theories provide a natural framework to understand the spontaneous 
breaking of parity and the origin of neutrino masses. In the classical left--right symmetric theories, the neutrino masses are 
generated through the type I and type II seesaw mechanisms~\cite{TypeI-2}. The type III seesaw mechanism can be nicely implemented in this context; see Ref.~\cite{FileviezPerez:2008sr}. For recent phenomenological studies of left--right symmetric models, see e.g., Refs.~\cite{LRPheno1,LRPheno2,Senjanovic:2010nq,LRPheno3}.

In general one cannot predict the scale where the left--right symmetry is realized, 
but one can think about the exciting possibility that this scale can be as low as a few TeV. In this case one can have spectacular signals 
at the Large Hadron Collider (LHC). Unfortunately, in this type of scenario one always has to postulate the existence of a large desert 
between the left--right scale and the high scale where one can understand the origin of higher-dimensional operators such as $ QQQL/\Lambda^2 $ 
mediating proton decay. As it is well known, the scale $\Lambda$ must be high, $\Lambda > 10^{14-16}$ GeV, in order to satisfy the experimental bounds on the proton decay lifetime.

In this paper, we investigate a new theoretical framework where the spontaneous breaking of parity is related to the spontaneous breaking of local baryon and lepton numbers.
This theory is based on the gauge group 
$SU(2)_L \otimes SU(2)_R \otimes U(1)_B \otimes U(1)_L,$
where $U(1)_B$ and $U(1)_L$ are the baryonic and leptonic symmetries.
In this context, there is no need to postulate the existence of a great desert in order to satisfy the experimental bounds on proton decay. On top of that we can understand 
the spontaneous breaking of $B$ and $L$ at the low scale. The idea of using this particular gauge group was studied in Ref.~\cite{He:1989mi}, where the authors investigated some solutions 
for the cancellation of anomalies. See also Ref.~\cite{Duerr} for a recent discussion of this type of theories.

Here, we show that the simplest anomaly-free theory based on the gauge group given above corresponds to the case where the neutrino masses are generated through 
the type III seesaw mechanism. We find that the spectrum for neutrinos is quite peculiar since one predicts the existence of two light sterile neutrinos without assuming any extra symmetries.
We briefly discuss the implications for collider physics.

\section{Theoretical Framework}
In order to define a simple theory where we can investigate the spontaneous breaking 
of parity, and baryon and lepton numbers, we use the gauge group
 \begin{equation*}
G_{LR}^{BL} =   SU(2)_L \otimes SU(2)_R \otimes U(1)_{B} \otimes U(1)_L \otimes {\cal{P}},
 \end{equation*}
 where ${\cal{P}}$ is the discrete left--right parity transformation. Here we do not display the QCD gauge group for simplicity. 
 The fields transform as
 \begin{align*}
  Q_L &= \begin{pmatrix}
          u_L \\ d_L
         \end{pmatrix}
         \sim \left(2,1,1/3,0\right),  
  Q_R  = \begin{pmatrix}
          u_R \\ d_R
         \end{pmatrix}
         \sim \left(1,2,1/3,0\right), \\
  \ell_L &= \begin{pmatrix}
          \nu_L \\ e_L
         \end{pmatrix}
         \sim \left(2,1,0,1\right), 
  \ell_R = \begin{pmatrix}
          \nu_R \\ e_R
         \end{pmatrix}
         \sim \left(1,2,0,1\right).
 \end{align*} 
As in any left--right symmetric theory, under the discrete left--right parity the fields transform as
 \begin{equation}
  Q_L \stackrel{\cal{P}}{\longleftrightarrow} Q_R \text{ and } \ell_L \stackrel{\cal{P}}{\longleftrightarrow} \ell_R,
 \end{equation}
 and the electric charge is defined as
 \begin{equation}
  Q = T_{3L} + T_{3R} + \frac{B-L}{2},
 \end{equation}
where $T_{3L}$ and $T_{3R}$ are the isospin under $SU(2)_L$ and $SU(2)_R$, respectively. 

As one expects, without extra fields we cannot define an anomaly-free theory. 
Here, the nontrivial anomalies are  
 \begin{align}
  \mathcal{A}_1\left( SU(2)_L^2 \otimes U(1)_B \right) &= 3/2, \\
  \mathcal{A}_2\left( SU(2)_L^2 \otimes U(1)_L \right) &= 3/2, \\
  \mathcal{A}_3\left( SU(2)_R^2 \otimes U(1)_B \right) &= - 3/2, \\
  \mathcal{A}_4\left( SU(2)_R^2 \otimes U(1)_L \right) &= - 3/2.
 \end{align}
Therefore, one needs to add new degrees of freedom to cancel these anomalies. In order to simplify our 
main task, we include extra fields that do not feel the strong interactions since the anomaly between 
the QCD group and baryon number is zero, i.e., $ \mathcal{A}_5 \left( SU(3)_C^2 \otimes U(1)_B \right) = 0$. 
Here we opt for the simplest solution where one can also generate neutrino masses through the seesaw mechanism. 
It is easy to show that the fields 
 \begin{equation*}
  \rho_L \sim \left(3,1,-3/4,-3/4 \right) \text{ and }
  \rho_R \sim \left(1,3,- 3/4,-3/4 \right)
 \end{equation*}
can provide the needed extra contributions for anomaly cancellation. 
Notice that the anomalies
\begin{eqnarray} 
&&\mathcal{A}_6 \left( U(1)_B^2 \otimes U(1)_L \right), \mathcal{A}_7 \left( U(1)_L^2 \otimes U(1)_B \right), \nonumber \\
&& \mathcal{A}_8 \left( U(1)_B \right), \mathcal{A}_9 \left( U(1)_B^3\right), \mathcal{A}_{10} \left( U(1)_L \right), \nonumber
\end{eqnarray}
and
$\mathcal{A}_{11}  \left( U(1)_L^3 \right)$ cancel as well. It is interesting to point out that the leptoquarks needed for anomaly cancellation have the right quantum 
numbers to generate neutrino masses through the type III seesaw mechanism. See Ref.~\cite{FileviezPerez:2008sr} 
for the implementation of the type III seesaw mechanism in left--right symmetric models and  Refs.~\cite{Barr:1986hj,BS,P1} for the implementation in other theories. 
Of course, one can imagine different solutions to cancel the anomalies, but we stick to the simplest one which also allows us to generate masses for all fields. Therefore, one can say that we have defined the simplest left--right theory based on the gauge group $G_{LR}^{BL}$.
\section{Leptoquarks and Neutrino Masses}
As already mentioned before, one can use the fields $\rho_L$ and $\rho_R$ to generate neutrino masses through the type III seesaw mechanism. 
In order to realize this idea, one needs to define the Higgs sector using the interactions
 \begin{eqnarray}
  -\mathcal{L} &\supset&  \bar{\ell}_L \left(  Y_3  \Phi + Y_4  \tilde{\Phi} \right)  \ell_R \nonumber \\
 & + & \lambda_D \left( \ell_L^T C  i \sigma_2 \rho_L H_L \ + \  \ell_R^T  C i \sigma_2 \rho_R H_R \right) \nonumber \\
 &  + &  \lambda_\rho  \ \text{Tr}  \left( \rho_L^T C  \rho_L + \rho_R^T C  \rho_R \right) S_{BL} \ + \  \text{h.c.}, \label{eq:lagrangian}
 \end{eqnarray}
 where the needed Higgs fields are given by
 \begin{eqnarray}
  \Phi =  \begin{pmatrix}
           \phi_1^0 & \phi_2^+ \\
           \phi_1^- & \phi_2^0
          \end{pmatrix}
        &\sim & \left(2,2,0,0 \right), \\
   H_L^T = \left( h_L^+ \ h_L^0 \right) &\sim & \left(2,1, 3/4, -1/4 \right), \\
   H_R^T =  \left( h_R^+ \ h_R^0 \right)  &\sim & \left(1, 2, 3/4, -1/4 \right), \\
   S_{BL} &\sim&  \left( 1,1, 3/2, 3/2 \right),
 \end{eqnarray}
 and $\tilde{\Phi} = \sigma_2 \Phi^\ast \sigma_2$. Notice that the Higgs sector is quite simple. 
 Now, under the left--right parity the new fields transform as
 \begin{equation}
  \rho_L \stackrel{\cal{P}}{\longleftrightarrow} \rho_R, 
  \Phi  \stackrel{\cal{P}}{\longleftrightarrow} \Phi^\dagger,   H_L  \stackrel{\cal{P}}{\longleftrightarrow}  H_R.
 \end{equation}  
The fields $\rho_L$ and $\rho_R$ are given by
\begin{align}
 \rho_L &= \frac{1}{2}\begin{pmatrix}
                     \rho_L^0 & \sqrt{2} \rho_L^+ \\
                     \sqrt{2} \rho_L^- & -\rho_L^0 \\
                    \end{pmatrix}  \text{ and } 
 \rho_R = \frac{1}{2}\begin{pmatrix}
                     \rho_R^0 & \sqrt{2} \rho_R^+ \\
                     \sqrt{2} \rho_R^- & -\rho_R^0 \\
                    \end{pmatrix}.
\end{align}

For spontaneous symmetry breaking (SSB), the Higgs fields will obtain 
the vacuum expectation values (VEVs)
\begin{align}
 \langle H_L \rangle &= \begin{pmatrix}
                        0 \\ v_L / \sqrt{2}
                       \end{pmatrix}, 
 \langle H_R \rangle = \begin{pmatrix}
                        0 \\  v_R / \sqrt{2}
                       \end{pmatrix}, \\
 \langle \Phi \rangle &= \begin{pmatrix}
                        v_1 & 0 \\ 0 & v_2
                       \end{pmatrix}, 
 \langle S_{BL} \rangle = v_{BL} / \sqrt{2}.
\end{align}
The hierarchy of the VEVs, and therefore the symmetry breaking sequence, is determined by the constraints on the model. $v_R$ is giving mass to the right-handed gauge boson, which has to be heavy (beyond TeV scale). $H_L$ and $\Phi$ have $SU(2)_L$ quantum numbers, and will therefore participate in the electroweak symmetry breaking. The sum of their VEVs can therefore not be beyond the electroweak scale, and we have to fulfill
\begin{equation}
 v_1^2 + v_2^2 + \frac{1}{2} v_L^2 = (174\, \text{GeV})^2.
\end{equation}
Furthermore, because $\Phi$ directly gives mass to the charged leptons, a la SM Higgs, $v_L \ll v_1, v_2$. This directly leads to parity violation,
\begin{equation}
 v_L \ll v_R. 
\end{equation}
Finally, $v_{BL}$ gives mass to $\rho_L$ and $\rho_R$ [see Eq.~\eqref{eq:lagrangian}], which have to be several hundred GeV or larger. In our setup, we want to use a type III seesaw mechanism (see Fig.~2), such that we have to demand $v_{BL} \gg v_R$.

After SSB, we can integrate out $\rho_L^0$ and $\rho_R^0$ and obtain the relevant Lagrangian for the neutrino masses
\begin{equation}
 -\mathcal{L}_\nu = M_\nu^D \overline{\nu}_L \nu_R - \frac{1}{2} M_{\nu_L}^{III} \nu_L^T C \nu_L -\frac{1}{2} M_{\nu_R}^{III} \nu_R^T C \nu_R + \text{h.c.} 
\end{equation}
Here, the masses are given by
\begin{align}
 \left(M_\nu^D\right)^{ij} &= Y_3^{ij} v_1 + Y_4^{ij} v_2, \\
 \left(M_{\nu_L}^{III}\right)^{ij} &= \frac{\lambda_D^i \lambda_D^j v_L^2}{4 \sqrt{2}\lambda_\rho v_{BL}}, \\
 \left(M_{\nu_R}^{III}\right)^{ij} & = \frac{\lambda_D^i \lambda_D^j v_R^2}{4 \sqrt{2}\lambda_\rho v_{BL}}.
\end{align}
A diagrammatical presentation of these mass terms is given in Figs.~1 and 2. 
\begin{figure}[t]
 \includegraphics{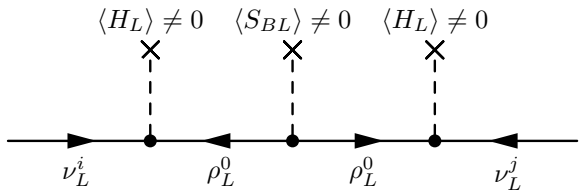}
 \caption{Type III seesaw for the left-handed neutrinos.}
\end{figure}
Notice that both mass terms $M_{\nu_L}^{III}$ and $M_{\nu_R}^{III}$ are generated through the type III seesaw mechanism and there is a simple relation between them:
\begin{equation}
M_{\nu_L}^{III}=\frac{v_L^2}{v_R^2} M_{\nu_R}^{III}.
\end{equation}

Now, parity violation, $v_L \ll v_R$, tells us that one must have the relation
\begin{equation*}
M_{\nu_L}^{III} \ll M_{\nu_R}^{III}.
\end{equation*}
This is the first consequence of having the type III seesaw mechanism in this context. 
The second consequence is that the Majorana mass matrix for the right-handed neutrinos given by $M_{\nu_R}^{III}$ has rank one.
Therefore, only one of the three right-handed neutrinos will have a non-zero Majorana mass. 
We can rotate the right-handed neutrinos such that $\nu_R \to U_R \nu_R$, and obtain
\begin{equation}
 -\mathcal{L}_\nu =  \tilde{M}_\nu^D \overline{\nu}_L \nu_R - \frac{1}{2} M_{\nu_L}^{III} \nu_L^T C \nu_L -\frac{1}{2} M_R \nu_{R}^{3 T} C \nu_R^3 + \text{h.c.} ,
\end{equation}
where  $\tilde{M}_\nu^D = M_\nu^D U_R$.
Then, $\nu_{R}^3$ will generate an additional Majorana mass for the left-handed neutrinos via the type I seesaw mechanism, such that we arrive at
\begin{equation}
 -\mathcal{L}_\nu = - \frac{1}{2} M_{\nu_L}^{LL} \nu_L^T C \nu_L + \left(\tilde{M}_\nu^D\right)^{i\alpha} \overline{\nu}_L^i \nu_R^{\alpha} + \text{h.c.},
\end{equation}
where $\alpha = 1,2$. The mass of the left-handed neutrinos is given by
\begin{equation}
 \left(M_{\nu_L}^{LL} \right)^{ij} = \left(M_{\nu_L}^{III}\right)^{ij} - \frac{1}{M_R} \left(\tilde{M}_\nu^D\right)^{i3} \left(\tilde{M}_\nu^D\right)^{j3}.
\end{equation}
Notice we can go to the basis where the matrix $M_{\nu_L}^{LL}$ is diagonal.
Therefore, the mass matrix for the light neutrinos in the theory can be written as
\begin{align}
 {\cal{M}}_{\nu}^{3+2} &= \begin{pmatrix}
                     0 & 0 & 0 & m_D^1 & m_D^2 \\
                     0 & m_1 & 0 & m_D^3 & m_D^4 \\
                     0 & 0 & m_2 & m_D^5 & m_D^6 \\
                     m_D^1 & m_D^3 & m_D^5 & 0 & 0 \\
                     m_D^2 & m_D^4 & m_D^6 & 0 & 0
\end{pmatrix} .
 \end{align}
This is a simple matrix which defines the mixing between the SM active neutrinos and the two extra sterile neutrinos.
In the limit $m_D^i \to 0$, the sterile neutrinos decouple and one of the active neutrinos is massless. 
The theory does not predict the numerical values of the coefficients in the above matrix, 
but one expects that the two extra sterile neutrinos can have 
mass below or at the eV scale. See Ref.~\cite{Kopp} for the constraints on sterile neutrinos with mass around the eV scale.
For other theories where one predicts the existence of light sterile neutrinos without assuming 
extra symmetries, see Refs.~\cite{MohapatranuR,BargernuR,GorannuR,LRnuR}.

In general one should investigate the possible impact of higher-dimensional operators to understand under 
which conditions the predictions at the renormalizable level is true. In this model one can write the higher-dimensional 
operator
\begin{equation}
 {\cal{O}}_{\nu_L}=c_L \ell_L \ell_L H_L H_L S_{BL}^\dagger/\Lambda^2,
\end{equation}
which generates neutrino masses of the order $M_{\nu_L} \sim v_L^2 v_{BL}/\Lambda^2$. Then, using the values $v_L \sim 1$ GeV 
and $v_{BL} \sim 10$ TeV, one needs $\Lambda \gtrsim 3 \times 10^3$ TeV to avoid a neutrino mass above 1 eV. The value of $v_L$ can of course be much smaller, such that this is a very naive bound on $\Lambda$. A similar 
operator can generate masses for the right-handed neutrinos, 
\begin{equation}
{\cal{O}}_{\nu_R}=c_R \ell_R \ell_R H_R H_R S_{BL}^\dagger/\Lambda^2.
\end{equation}
In this case the right-handed neutrino mass reads as $M_{\nu_R} \sim v_R^2 v_{BL}/\Lambda^2$. Using $v_R \sim 1$ TeV and $v_{BL} \sim 10$ TeV and the above bound for $\Lambda$, one gets $M_{\nu_R} < 1$ MeV. Notice that the two operators ${\cal{O}}_{\nu_R}$ 
and ${\cal{O}}_{\nu_L}$ are connected by the left--right parity. After spontaneous symmetry breaking these operators also induce baryon number violation, but they do not induce visible baryon number violating decays in the quark sector such as proton decay. There are higher-dimensional operators that induce baryon number violation in the quark sector, e.g.,
\begin{equation}
 \mathcal{O}_{20} = \frac{1}{\Lambda^{16}} \left( Q_L Q_L Q_L \ell_L \right)^3 S_{BL}^\dagger  S_{BL}^\dagger,
\end{equation}
which, however, is of dimension 20 and therefore strongly suppressed. Hence, there is no need to postulate a large desert in this scenario.

\begin{figure}[t]
 \includegraphics{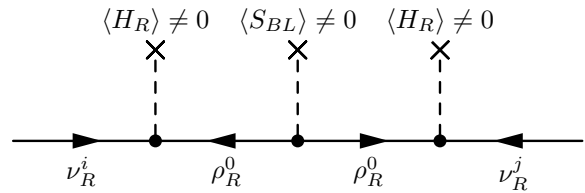}
 \caption{Type III seesaw for the right-handed neutrinos.}
\end{figure}

Recently, the Planck Collaboration~\cite{Ade:2013zuv} has set limits on extra relativistic degrees
of freedom. The model studied in this Letter predicts the possible existence of extra light degrees of 
freedom and one should investigate the possible cosmological constraints. This study is beyond the scope 
of this paper, but we would like to mention how a consistent picture can be achieved. 
The contribution of the extra neutrinos to $N_\text{eff}$ depends on the mass of the new gauge bosons, 
$Z_1, Z_2$, and  $W_R$, since they can be in thermal equilibrium with the SM plasma through their 
interactions. In Refs.~\cite{Concha,Anchordoqui:2012qu} the authors have investigated the constraints 
on the new forces which keep the right-handed neutrinos in thermal equilibrium. They have shown 
that if these gauge bosons are in the TeV region one does not have a large contribution to the 
effective number of relativistic degrees of freedom because the neutrinos decouple very early.
The LHC bounds on the mass of the new gauge bosons are severe and one can satisfy the cosmological constraints at the same time.
\section{Summary}
We have investigated the possibility of defining a simple theory for the spontaneous breaking of parity, and baryon and lepton numbers at the low scale. 
We have found that the needed leptoquark fields for anomaly cancellation are the same fields 
that generate masses for the left-handed and right-handed neutrinos through 
the type III seesaw mechanism. This theory can be considered as the simplest theory 
with these features. The local baryonic and leptonic symmetries can be broken 
at the low scale and we do generate any interactions mediating proton decay.

The spectrum for neutrinos in this theory has been studied in detail, showing that one 
predicts the existence of two light right-handed neutrinos. The possible impact of 
the higher-dimensional operators has been mentioned. The existence of these 
light degrees of freedom can change the phenomenology 
of the new gauge bosons at colliders. In this case the neutral gauge bosons 
can decay into the extra right-handed light neutrinos, $Z_i \to \bar{\nu}_R \nu_R$, 
and more importantly, the decays $W_R \to \bar{\nu}_R e_R$ are allowed. 
These decays can change the present collider bounds on the $W_R$ mass based 
on the lepton number violating decays of the right handed neutrinos. 
All these aspects will be investigated in a future publication.

{\textit{Acknowledgments}}:
{\small{P. F. P. thanks M. B. Wise for discussions. 
M.\ D.\ is supported by the IMPRS-PTFS of the Max Planck Society.}

\end{document}